\documentstyle[twoside]{article}

\catcode`\@=11
\long\def\@makefntext#1{
\protect\noindent \hbox to 3.2pt {\hskip-.9pt  

$^{{\eightrm\@thefnmark}}$\hfil}#1\hfill}               

\def\thefootnote{\fnsymbol{footnote}}
\def\@makefnmark{\hbox to 0pt{$^{\@thefnmark}$\hss}}    

\def\ps@myheadings{\let\@mkboth\@gobbletwo
\def\@oddhead{\hbox{}
\rightmark\hfil\eightrm\thepage}   

\def\@oddfoot{}\def\@evenhead{\eightrm\thepage\hfil
\leftmark\hbox{}}\def\@evenfoot{}
\def\sectionmark##1{}\def\subsectionmark##1{}}

\oddsidemargin=\evensidemargin
\renewcommand{\thefootnote}{\fnsymbol{footnote}}

\newcounter{sectionc}\newcounter{subsectionc}\newcounter{subsubsectionc}
\renewcommand{\section}[1] {\vspace{12pt}\addtocounter{sectionc}{1} 

\setcounter{subsectionc}{0}\setcounter{subsubsectionc}{0}\noindent 

        {\tenbf\thesectionc. #1}\par\vspace{5pt}}
\renewcommand{\subsection}[1] {\vspace{12pt}\addtocounter{subsectionc}{1} 

        \setcounter{subsubsectionc}{0}\noindent 

        {\bf\thesectionc.\thesubsectionc. {\kern1pt \bfit #1}}\par\vspace{5pt}}
\renewcommand{\subsubsection}[1] {\vspace{12pt}\addtocounter{subsubsectionc}{1}
        \noindent{\tenrm\thesectionc.\thesubsectionc.\thesubsubsectionc.
        {\kern1pt \tenit #1}}\par\vspace{5pt}}

\newcounter{appendixc}
\newcounter{subappendixc}[appendixc]
\newcounter{subsubappendixc}[subappendixc]
\renewcommand{\thesubappendixc}{\Alph{appendixc}.\arabic{subappendixc}}
\renewcommand{\thesubsubappendixc}
        {\Alph{appendixc}.\arabic{subappendixc}.\arabic{subsubappendixc}}

\renewcommand{\appendix}[1] {\vspace{12pt}
        \refstepcounter{appendixc}
        \setcounter{figure}{0}
        \setcounter{table}{0}
        \setcounter{lemma}{0}
        \setcounter{theorem}{0}
        \setcounter{corollary}{0}
        \setcounter{definition}{0}
        \setcounter{equation}{0}
        \renewcommand{\thefigure}{\Alph{appendixc}.\arabic{figure}}
        \renewcommand{\thetable}{\Alph{appendixc}.\arabic{table}}
        \renewcommand{\theappendixc}{\Alph{appendixc}}
        \renewcommand{\thelemma}{\Alph{appendixc}.\arabic{lemma}}
        \renewcommand{\thetheorem}{\Alph{appendixc}.\arabic{theorem}}
        \renewcommand{\thedefinition}{\Alph{appendixc}.\arabic{definition}}
        \renewcommand{\thecorollary}{\Alph{appendixc}.\arabic{corollary}}
        \renewcommand{\theequation}{\Alph{appendixc}.\arabic{equation}}
        \noindent{\tenbf Appendix \theappendixc #1}\par\vspace{5pt}}
\newcommand{\subappendix}[1] {\vspace{12pt}
        \refstepcounter{subappendixc}
        \noindent{\bf Appendix \thesubappendixc. {\kern1pt \bfit #1}}
        \par\vspace{5pt}}
\newcommand{\subsubappendix}[1] {\vspace{12pt}
        \refstepcounter{subsubappendixc}
        \noindent{\rm Appendix \thesubsubappendixc. {\kern1pt \tenit #1}}
        \par\vspace{5pt}}

\newcommand{\textlineskip}{\baselineskip=13pt}
\newcommand{\smalllineskip}{\baselineskip=10pt}

\def\eightcirc{
\begin{picture}(0,0)
\put(4.4,1.8){\circle{6.5}}
\end{picture}}
\def\eightcopyright{\eightcirc\kern2.7pt\hbox{\eightrm c}}

\newcommand{\copyrightheading}[1]
        {\vspace*{-2.5cm}\smalllineskip{\flushleft
        {\footnotesize International Journal of Modern Physics A, #1}\\
        {\footnotesize $\eightcopyright$\, World Scientific Publishing
         Company}\\
         }}

\newcommand{\publisher}[2]{{\begin{center}\footnotesize\smalllineskip 

        Received #1\\
        \end{center}
        }}

\def\abstracts#1#2#3{{
        \centering{\begin{minipage}{4.5in}\baselineskip=10pt\footnotesize
        \parindent=0pt #1\par 

        \parindent=15pt #2\par
        \parindent=15pt #3
        \end{minipage}}\par}}

\renewenvironment{thebibliography}[1]
        {\frenchspacing
         \ninerm\baselineskip=11pt
         \begin{list}{\arabic{enumi}.}
        {\usecounter{enumi}\setlength{\parsep}{0pt}
         \setlength{\leftmargin 12.7pt}{\rightmargin 0pt} 
         \setlength{\itemsep}{0pt} \settowidth
        {\labelwidth}{#1.}\sloppy}}{\end{list}}

\newcounter{itemlistc}
\newcounter{romanlistc}
\newcounter{alphlistc}
\newcounter{arabiclistc}

\newcommand{\fcaption}[1]{
        \refstepcounter{figure}
        \setbox\@tempboxa = \hbox{\footnotesize Fig.~\thefigure. #1}
        \ifdim \wd\@tempboxa > 5in
           {\begin{center}
        \parbox{5in}{\footnotesize\smalllineskip Fig.~\thefigure. #1}
            \end{center}}
        \else
             {\begin{center}
             {\footnotesize Fig.~\thefigure. #1}
              \end{center}}
        \fi}

\newcommand{\tcaption}[1]{
        \refstepcounter{table}
        \setbox\@tempboxa = \hbox{\footnotesize Table~\thetable. #1}
        \ifdim \wd\@tempboxa > 5in
           {\begin{center}
        \parbox{5in}{\footnotesize\smalllineskip Table~\thetable. #1}
            \end{center}}
        \else
             {\begin{center}
             {\footnotesize Table~\thetable. #1}
              \end{center}}
        \fi}

\def\@citex[#1]#2{\if@filesw\immediate\write\@auxout
        {\string\citation{#2}}\fi
\def\@citea{}\@cite{\@for\@citeb:=#2\do
        {\@citea\def\@citea{,}\@ifundefined
        {b@\@citeb}{{\bf ?}\@warning
        {Citation `\@citeb' on page \thepage \space undefined}}
        {\csname b@\@citeb\endcsname}}}{#1}}

\newif\if@cghi
\def\cite{\@cghitrue\@ifnextchar [{\@tempswatrue
        \@citex}{\@tempswafalse\@citex[]}}
\def\citelow{\@cghifalse\@ifnextchar [{\@tempswatrue
        \@citex}{\@tempswafalse\@citex[]}}
\def\@cite#1#2{{$\null^{#1}$\if@tempswa\typeout
        {IJCGA warning: optional citation argument 

        ignored: `#2'} \fi}}

\def\pmb#1{\setbox0=\hbox{#1}
        \kern-.025em\copy0\kern-\wd0
        \kern.05em\copy0\kern-\wd0
        \kern-.025em\raise.0433em\box0}

\def\fnt#1#2{\footnotetext{\kern-.3em
        {$^{\mbox{\scriptsize #1}}$}{#2}}}

\def\fpage#1{\begingroup
\voffset=.3in
\thispagestyle{empty}\begin{table}[b]\centerline{\footnotesize #1}
        \end{table}\endgroup}

\def\runninghead#1#2{\pagestyle{myheadings}
\markboth{{\protect\footnotesize\it{\quad #1}}\hfill}
{\hfill{\protect\footnotesize\it{#2\quad}}}}
\font\tenrm=cmr10
\font\tenit=cmti10 

\font\tenbf=cmbx10
\font\bfit=cmbxti10 at 10pt
\font\ninerm=cmr9

\font\eightrm=cmr8

\def\qed{\hbox{${\vcenter{\vbox{                        
   \hrule height 0.4pt\hbox{\vrule width 0.4pt height 6pt
   \kern5pt\vrule width 0.4pt}\hrule height 0.4pt}}}$}}

\def\half{{1\over2}}
\def\c#1{{\cal #1}}
\def\IR{\relax{\rm I\kern-.18em R}}

\def\corru#1#2{\langle #2\rangle_{{}_{#1}}}
\def\prodone#1{\prod_{j=1}^{#1}}

\def\al#1{a_{#1}}

\renewcommand{\thefootnote}{\fnsymbol{footnote}}        

\begin{document}

\runninghead{High Energy Scattering of Non-Critical Strings}  
{High Energy Scattering of Non-Critical Strings}

\normalsize\textlineskip
\thispagestyle{empty}
\setcounter{page}{1}

\copyrightheading{}                     

\vspace*{0.88truein}

\fpage{1}
\centerline{\bf HIGH ENERGY SCATTERING OF NON-CRITICAL STRINGS} 
\vspace*{0.37truein}
\centerline{\footnotesize KENICHIRO AOKI}
\vspace*{0.015truein}
\centerline{\footnotesize\it Department of Physics, Hiyoshi Campus, Keio 
University, Kouhoku-ku}
\baselineskip=10pt
\centerline{\footnotesize\it Yokohama 223, Japan}
\vspace*{10pt}
\centerline{\footnotesize ERIC D'HOKER}
\vspace*{0.015truein}
\centerline{\footnotesize\it Department of Physics, University of California}
\baselineskip=10pt
\centerline{\footnotesize\it Los Angeles, CA 90024, USA}
\vspace*{0.225truein}
\publisher{27 September 1996}

\vspace*{0.21truein}
\abstracts{We summarize recent work, in which we consider scattering
amplitudes of non-critical strings in the limit where the
energy of all external states is large compared to the string tension.
We show that the high energy limit is dominated by a saddle point
that can be mapped onto  an electrostatic equilibrium configuration
of an assembly of charges associated with the external states, together
with a density of charges arising from the Liouville field. The
Liouville charges accumulate on line segments, which produce 
quadratic branch cuts on the worldsheet. The electrostatics
problem is solved for string tree level in terms of hyperelliptic
integrals and is given explicitly for the 3- and 4-point functions.
For generic values of the central charge, the high energy limit 
behaves in a string-like fashion, with exponential energy dependence.}{}{}

\vspace*{1pt}\textlineskip
\section{Introduction}  
\vspace*{-0.5pt}
\noindent
In this lecture, we shall summarize our recent work$^1$ on the high 
energy behavior of non-critical bosonic strings. Detailed derivations,
as well as extended references to the literature, may be found in [1].

While critical string theory is the prime candidate for a unified
theory of matter and gravity, non-critical string theory may be of
most interest precisely there where gravity is of no concern.
Examples of this situation arise in statistical mechanics systems on random
lattices, in Polyakov's mapping of the three dimensional Ising model
onto a three dimensional fermionic string, and perhaps even in string
type reformulations of Quantum Chromodynamics. All of these physical
systems admit some reformulation in terms of non-critical string theory.

Much has been accomplished in the study of non-critical string theory
(for reviews on the subject, see e.g. [2]).
Our understanding of non-critical string theory to date is particularly
advanced for bosonic models with rational central charge $c<1$. 
The mapping between discretized random surfaces and random matrices
combined with the double scaling limit produces exact results for
the correlation functions to all orders in perturbation theory.

Surprisingly, it has turned out to be very difficult to reproduce, with
the help of the Liouville approach to non-critical string theory, even the
simplest results obtained via matrix models. For example, the evaluation of 
scattering amplitudes to tree level presents serious obstacles, 
which have not, in general, been overcome to date. Despite these obstacles, 
the Liouville approach to non-critical string theory is of great value. 
This is especially so since a direct reformulation of fermionic non-critical 
strings in terms of matrix models is not available, while the generalization 
from the bosonic to the fermionic string is essentially straightforward within 
the Liouville approach.
Also, the applicability of the matrix model approach seems to be limited to 
$c\leq 1$ while the Liouville approach again allows for a passage into the 
$c>1$ domain.  
One of the most basic obstacles to reaching beyond the $c=1$
barrier (within the Liouville field theory approach), is the
appearance of conformal primary fields with complex weights, and
thus of string states with complex masses. In a series of 
ingenious papers [3], it was proposed that
the string spectrum may be restricted to a subset of ``physical
states'', that have real conformal weights only. This restriction 
appears to be possible only provided 
space-time dimension assumes certain special values : 1, 7, 13 and 19
for the bosonic string, and 1, 3, 5 and 7 for the fermionic string.

In our work,$^1$ we developed calculational methods
that allow for an evaluation of scattering amplitudes in non-critical 
string theory. We proposed an evaluation of non-critical string
amplitudes for any complex $c$, in the limit where the energies
of incoming and outgoing string states are all large compared to
the (square root of the) string tension. 
We showed that the Liouville approach lends itself naturally 
to taking the high energy limit, where the integral 
representations for the amplitudes become tractable, for any
complex value of $c$. To string tree level, we succeed in 
producing explicit formulas for the limit in terms of 
hyper-elliptic integrals. We shall not, at this stage, 
perform any truncation on the spectrum of states in the 
non-critical string theory. Thus, our results are applicable
to non-critical string theories in general, including 
those in which the Liouville field is reinterpreted as
an extra dimension of space-time.

For string theory in the critical dimension, the high energy
limit of scattering amplitudes is dominated by a saddle point in
the positions of the vertex operators for external string
states, as well as in the moduli of the surface.  This problem
is equivalent to finding the equilibrium configuration of an
array of electrostatic Minkowskian charges (attached to the
vertex operators) on a surface of variable shape. In a series 
of beautiful papers$^4$, it was shown how the
saddle point can be constructed by symmetry arguments, for the
four point function, to any order in perturbation theory.

For non-critical string theory, the high energy limit is still
dominated by a saddle point, which is equivalent to the equilibrium
configuration of an array of (complex) charges on a surface of
variable shape. In addition to the charges from the external
vertex operators, we now also have charges from the Liouville
exponential operator. In fact, the number of Liouville charges
on the surface increases linearly with energy and, in the limit
of large energy, accumulate onto a continuous charge density. We
shall show that this Liouville charge density consists of line
segments, producing quadratic branch cuts on the worldsheet.

We shall solve explicitly the equivalent electrostatics problem
for a worldsheet with the topology of a sphere (tree level) in
terms of hyper-elliptic functions, and use it to deduce the high
energy limit of tree level scattering amplitudes.  The solution
in this limit is valid for any complex value of the matter central charge
$c$, and we use analytic continuation to define the non-critical
string amplitudes throughout the complex $c$ plane. For higher
genus topologies, the solution involves quadratic branch cuts of
higher genus surfaces, but we shall postpone a full derivation
of this case to a later publication.

The main physical result is that, at least for generic values of the 
matter central charge $c$, the non-critical amplitudes behave
in a string like fashion, with exponential dependence on the
energy scale, in the limit of high energy. While it is logically
possible that this generic exponential behavior could be
absent (and replaced by power-like behavior)
at isolated points in the complex $c$ plane, we believe
that this is unlikely to occur in the region $1<c<25$.
It is thus unlikely that the non-critical string theories
in this region ever become ``quantum field theories''.

Much work remains to be done: Our method for obtaining the
correlation functions generalizes to loop amplitudes in
non--critical string theory.  In this case, the problem reduces
to a two dimensional electrostatics problem on  double
coverings of higher genus Riemann surfaces.  Also, the continuum
approach naturally extends to the supersymmetric case.  Perhaps
more importantly, the correlation functions may be used to
analyze the physics of non-critical string theories.  Given the
diverse applications of non--critical string theory, this is of
great interest.  These points are currently under investigation.
\textheight=7.8truein
\setcounter{footnote}{0}
\renewcommand{\thefootnote}{\alph{footnote}}
\section{Liouville Field Theory Approach to Non-Critical Strings}\noindent

\noindent
The basic ingredients in the Liouville field
theory formulation of bosonic non-critical string theory are
as follows. The starting point is a ``matter'' conformal field theory,
with central charge $c$, and fields $x^\mu (z)$, $\mu =1.\cdots, c$, describing Poincar\'e invariant string dynamics in a
$c$-dimensional space-time. 

The above conformal field theory is coupled to a quantized
worldsheet metric $g$, which, in conformal gauge, decomposes
into the Liouville field $\phi (z)$ and a fiducial metric $\hat
g(m_j)$ that only depends on the moduli of the Riemann
surface $\Sigma$, with $g=\hat g \exp\{2\phi\}$.  The action for
the Liouville field is
\begin{equation}
  S_{L} =\frac{1}{4 \pi} \int\sqrt{\hat g} \left[
  \half \phi \Delta _{\hat g} \phi -\kappa  R_{\hat g} \phi
  +{ \mu }  e^{ \alpha\phi}        \right]
\end{equation}
Here $R_{\hat g}$ is the Gaussian curvature of the metric 
$\hat g$, and the coupling constants $\kappa$ and 
$\alpha$ are given in terms 
of the matter central charge $c$ as follows $ 3\kappa ^2 = 25-c$.  
The gravitationally dressed vertex operators
may be obtained as follows
\begin{equation}
  {\cal V}_\delta \equiv \int d^2z {\cal P} (\partial x^\mu)
  e^{ik\cdot x + \beta\phi(z)},\qquad 
  \beta(\delta ) ={-\sqrt{25-c}+\sqrt{1-c+24\delta
      }\over2\sqrt3} 
\end{equation}
Here, ${\cal P}$ is a polynomial in the derivatives of $x^\mu$ 
of degree $\Delta$, so that $\delta = \Delta + \frac{1}{2} k^2$.
For simplicity, we shall assume that ${\cal P}$ is independent
of $\phi$.
The coupling constant $\alpha$ is given by $\alpha = \beta (0)$, and
the analytic continuation in $c$ and in the external momentum
dictates which branches of the square roots should be chosen.

We shall now compute out the correlation function for $N$ external
gravitationally dressed vertex operators.
The correlation functions of the matter part are standard, so we
concentrate on the evaluation of the Liouville correlation functions.
To evaluate the Liouville correlation functions, we follow the
procedure of [5].
We split the Liouville field $\phi$ as follows $\phi = \phi _0 +
\varphi$, where $\phi _0$ is constant on the worldsheet and
$\varphi$ is orthogonal to constants. The integration splits
accordingly, and the integral over $\phi _0$ may be carried out
as follows
\begin{equation}
  \int D_{\hat g}\phi\,e^{-S_L}\prodone N e^{\beta_j\phi(z_j)} =
  { \Gamma(-s) \mu ^s \over \alpha (4\pi)^s}
  \int D_{\hat g}\varphi \ e^{-S'_L}\left(\int\!\!\sqrt{\hat g}\
    e^{\alpha\varphi}\right)^s
  \prodone N e^{\beta_j\varphi(z_j)}
\end{equation}
The new Liouville action $S_L'$ is obtained by setting $\mu =0$, and
the variable $s$ is a scaling dimension, defined by $\alpha$, $\kappa$,
$\beta _j$ and the genus of the worldsheet $h$ as follows
\begin{equation}
\alpha s=-{\kappa}(1-h)-\sum_{j=1}^N\beta_j
\end{equation}
In general, $s$ does not have to be integer, not even
rational. The prescription of [5] is to 
proceed and carry out the functional integration over $\varphi$
as if $s$ were an integer, and then later on continue in $s$. We
are confident that this procedure is reliable in view of the
semi-classical analysis carried out in the second reference of [5]. 
\section{The High Energy Limit as a Problem in Electrostatics}
\noindent
To tree level, the worldsheet topology is that of the sphere (or
by stereographic projection, of the complex plane), there are no
moduli, and all determinant factors are constants. The
Green function is the electrostatic potential on the two
dimensional plane, given by $ G(z,z')= -\ln  |z-z'|^2$. 
Tree-level non-critical amplitudes --- evaluated for vertex
operators that are exponentials only --- then reduce to a simple
multiple integral expression
\begin{equation}
    \corru{}{\prod_{i=1}^N \c V _i}
    =  
    {\Gamma(-s) \mu ^s \over\alpha (4\pi )^s}
    \int  \prod_{i=1}^Nd^2z_i 
    \prod_{ i<j\atop i,j=1}^N \left|z_i-z_j\right|^{2u_{ij}} 
    \int \prod_{p=1}^sd^2w_p
    \left|z_j-w_p  \right|^{-2\alpha\beta_j }
    \prod_{p,q=1\atop p<q}^s\left|w_p-w_q\right|^{-2\alpha^2 }
    \label{eq:corrfn}
\end{equation}
where $u_{ij} = -\beta _i \beta _j +k_i \cdot k_j$.  
The problem of evaluating correlation functions in non-critical
string theory is seemingly reduced to the problem of computing a
finite dimensional multiple integral with respect to 
$z_i, \ w_a$ over the complex plane.  
Since $s$ is not, in general, an integer however, the
correlation function is not well defined as it stands. 
{}From the arguments presented in [5], it is clear that the 
original expression for the amplitudes  
is complex analytic in the external momenta and in the 
central charge $c$, even though the intermediate expressions 
only make sense for integer $s$. 
Thus, the results obtained for integer $s$ will have to be 
analytically continued in $s$, which can be achieved through 
a combination of analytic continuation 
in the external momenta (just as in the critical string)
and in the central charge $c$.

For rational $c<1$, and to string tree level, it was proposed in
[5] to analytically continue in the variable $s$, using certain
rearrangement formulas for ratios of Euler $\Gamma$-functions
(that are specific to tree level). The validity of this
procedure is justified, after the fact, since it produces
agreement with results from matrix models. More importantly,
agreement can be established from first principles, as was 
shown in the second reference in [5], using a saddle point 
approximation in the limit 
when $\alpha\rightarrow0$, i.e. when $c\rightarrow \infty$.
We shall take these analyticity properties as a definition
for the amplitudes away from $c<1$ and rational.

For tree level amplitudes, the above multiple integrals are of
the same type as those discussed by Selberg and in [6].
The 3-point function was obtained in their work for
arbitrary parameters, but results on the 4-point function are
limited to $c<1$ conformal matter. In general, these integrals are not
available in explicit form.

We propose to evaluate the non-critical string correlation
functions in the limit where the energies and momenta of the
external string states all become large compared to the square
root of the string tension. We shall define the high energy
limit by rescaling all momenta $k_i$ by a common factor $\lambda
\rightarrow \infty$. 
We have the asymptotic behavior 
$ k_i  \rightarrow  \lambda k_i,\   \beta _i \rightarrow 
  \pm \lambda |k_i |  +\c O(1) $.
The scaling properties of other quantities are easily deduced
from the above~: $u_{ij} $ scales like $\lambda ^2$, $s$ scales
like $\lambda ^1$ while $c$ and $\alpha$ scale like $\lambda
^0$. Notice that external vertex operators always remain
conformally invariant under this scaling.

To determine the high energy limit of the non-critical
scattering amplitude, we begin by analyzing the
high energy behavior of the electrostatic energy function
 $\c E_0$ corresponding to the correlation function of Eq.
(\ref{eq:corrfn}), defined by
\begin{equation}
\c E_0(z_i, w_p) 
=  
    - \sum _{i,j=1\atop i<j}^N u_{ij}  \ln |z_i-z_j|^2 
    +\sum _{j=1}^N\sum _{p=1}^s \alpha\beta_j \ln |z_j-w_p |^2
    +\sum_{p,q=1\atop p<q}^s \alpha ^2 \ln |w_p-w_q|^2 
\end{equation}
{}From the
expression for $\c E_0$, it can be readily shown that $\c E_0$
scales like $\lambda ^2$ for large $\lambda$.  This is manifest
for the first term in $\c E_0$, but the next two terms also
scale like $\lambda ^2$ for large $\lambda$. Although the
couplings in the second term only scale linearly in $\lambda$,
the number of Liouville insertion points, $s$, grows like
$\lambda$. In the third term, the coupling $\alpha$ scales like
$\lambda ^0=1$, but there are now $s^2$ Liouville insertion
points, so again this term scales like $\lambda ^2$.

The next ingredient needed in the determination of the high
energy limit of the non-critical scattering amplitudes is the
degree of dependence of this limit on any specific matter
conformal field theory. The most important simplification in
this respect comes from the observation that the conformal
primary fields $\c P _\Delta$ involve and produce only
polynomial dependence on the space-time momenta $k_i$. Thus, in
the high energy limit, where the contributions from the saddle
point will be generically exponential (as we shall establish
below), we may neglect the polynomial contributions from the
vertex functions $\c P _\Delta$. Thus, only the exponential
vertex operator parts contribute to the high energy limit.

The saddle point equations for the integral are just the
equations for electrostatic equilibrium of the associated
electrostatics problem. They are given by
\begin{equation}
    -\sum_{j=1\atop j\not=i}^N{2b_{ij}\over z_i-z_j}
    +{1 \over s} \sum_{p=1}^s {a_i \over z_i-w_p}  =  0
    \qquad \qquad
   {1\over s}\sum_{q\not=p\atop q=1}^s{2\over
      w_p-w_q}+\sum_{j=1}^N{\al j\over w_p-z_j} =  0
\label{eq:elec}
\end{equation}
Here, we have defined parameters
$\al i\equiv 2\beta_i/(\alpha s)$ and $b_{ij} \equiv u_{ij} /
(\alpha s)^2$ both of which scale like $\lambda ^0$ in the limit
of large $\lambda$. Also, each summation over the number of
Liouville charges at $w_p$, $p=1,\cdots ,s$ has been divided by a
factor of $s$, so that the entire equations  scale
like $\lambda ^0$ in the limit of large $\lambda$.

The variables $\bar z _i$ and $\bar w _p$ satisfy
the same equation (\ref{eq:elec}), 
with $z_i$ and $w _p$ replaced by $\bar z_i$ and
$\bar w_p$ respectively. When the charges $a_i$ and $b_{ij}$
are real, those respective equations are just the complex
conjugates of one another. But when the charges $a_i$ and
$b_{ij}$ are taken to be complex, the equations are no longer
complex conjugates of one another, and $\bar z_i$ and $\bar
w_p$ at the saddle point are no longer the complex conjugates
of $z_i$ and $w_p$ respectively. 

 We wish to solve the equations (\ref{eq:elec}) in the limit where
$s \rightarrow \infty$, while keeping $a_i$, $b_{ij}$ and the
number of vertex charges $N$ fixed. The most difficult part of
this problem is the solution of the second equation, 
for the following reasons. Since the number of
Liouville charges at $w_p$ tends to $\infty$, they must 
accumulate somewhere, possibly at infinity. 
A priori, the limiting 
distribution might correspond to two-dimensional regions of charge, 
to one-dimensional line segments, to isolated points, or even to more 
exotic arrangements of fractional dimension such as Cantor sets.
 
We shall start by providing an answer to this question first, 
by carefully keeping the Liouville charges at $w_p$
isolated, and taking the limit only when completely safe.
In [1], we showed that this problem can be solved exactly
in the case of the three point function with $N=3$, and we 
found there that the Liouville charges accumulate onto 
a single line segment. More generally, we shall find that the 
Liouville charges accumulate onto a collection of $N-2$ curve 
segments. 
\section{Solution of the Electrostatics Problem to Tree Level}
\noindent
To study equation (\ref{eq:elec}) for general $N$, we make use
of a complex potential $W(z)$ for the charges $z_i$ and
re-express the equation (\ref{eq:elec}) for the Liouville charges
at $w_p$ as
\begin{equation}
  \frac{1}{s} \sum_{q=1\atop q\not =p}^s{1\over
    w_p-w_q}=\half W'(w_p) 
  \ \ \hbox{where}\ \ \ 
  W(z)\equiv -\sum_{j=1}^{N-1}\al j\ln( z_j-z)
\label{eq:msaddle}
\end{equation}
This equation, for general $W(z)$, is just the electrostatics
condition for an assembly of $s$ charges in the presence of an
external potential $W(z)$ --- given here by the potential
generated by the charges at $z_i$. We also introduce a complex
analytic generating function $\omega (z)$, defined by
\begin{equation}
  \omega(z)={1\over s}\sum_{p=1} ^s {1\over w_p-z}
\label{eq:omdef}
\end{equation}
which, physically, is just the electric field produced by the
Liouville charges at $w_p$. Its divergence is obtained by
applying the Cauchy-Riemann operator, and yields the
(two-dimensional) electric charge density with unit integral
over the plane.

One may re-express the set of $s$ equations (\ref{eq:msaddle})
in terms of the following Riccati equation for $\omega(z)$ by
introducing an auxiliary potential $R(z)$
\begin{equation}
  \omega^2(z)-{1\over s}\omega'(z)+
  W'(z)\omega(z)+ {1 \over 4}R(z)=0,\qquad
  R(z) \equiv {4\over s}\sum_{p=1}^s{W'(w_p)-W'(z)\over
   w_p-z}
\label{eq:omegaeq}
\end{equation}
For general $W(z)$, it would not be possible to carry out the
sum in the definition of $R(z)$ in any simple way. When $W(z)$
is a rational function of $z$ however, as is the case here,
$R(z)$ is also rational, with poles at precisely the same
locations as $W(z)$:
\begin{equation}
  R(z)=\sum_{i=1}^{N-1}{R_i\over z-z_i},\qquad{\rm where}\qquad
  R_i = {4\al i \over s}\sum _{p=1}^s {1 \over w_p -z_i}
\label{eq:rsol}
\end{equation}
The fact that we have been able to determine the functional form
of $R(z)$ explicitly, in terms of a finite number of parameters
is perhaps the most important ingredient in our solution of the
associated electrostatics problem.

All that precedes is still an exact transcription of the
electrostatics equations (\ref{eq:elec}), valid for any finite
number of Liouville charges at $w_p$. We shall now bring about
one further simplification by using the approximation in which
the number of charges $s$ is large. (Recall that, in the
original non-critical string problem, this limit corresponds to
high energy of all external string states.)

The potential $W(z)$ is independent of $s$, while the electric
field $\omega(z)$ and the auxiliary potential $R(z)$ converge to
finite limits as $s\rightarrow \infty$. Thus, the electrostatics
equation for $\omega (z)$ of ~(\ref{eq:omegaeq}) can be
simplified in this limit, as the term in $\omega '(z)$ is
suppressed by a factor of $1/s$ and may be dropped. Instead of
the Riccati equation of (\ref{eq:omegaeq}), we obtain
a quadratic equation which is easily solved. 
We obtain 
\begin{equation}
  \omega(z)={1\over2}\left[-W'(z)\pm\sqrt{W'^2(z)-R(z)}\right],
\label{eq:omsol}
\end{equation}
The sign in front of the square root should be chosen so that
the poles in $\omega (z)$, located at the points $z_i$, are
absent when the charges $a_i$ are all real and positive.
Eq. (\ref{eq:omsol}) immediately determines the density of
Liouville charges; since the solution $\omega (z)$ in
(\ref{eq:omsol}) is holomorphic away from the possible poles at
$z_i$, and away from the quadratic branch cuts arising from the
square root, we see that the Liouville charge accumulate on the
branch cuts of $\omega$.

The positions of the associated branch points are 
most easily exhibited by recasting the solution for $\omega(z)$ 
as follows~:
\begin{equation}
W'(z)^2 - R(z) = Q_{2N-4} (z)  \prod _{i=1} ^{N-1} (z-z_i)^{-2}
\label{eq:polyn}
\end{equation}
Here, $Q_{2N-4}(z)$ is a polynomial in $z$, which is of degree
$2N-4$, in view of the fact that the sum of all $R_i$ vanishes. 
We find
\begin{equation}
Q_{2N-4} (z) = (a+2) ^2 \prod _{k=1} ^{2N-4} (z-x_k) 
\qquad\qquad
a = \sum _{i=1} ^{N-1} a_i
\label{eq:polynorm}
\end{equation}
The function $\omega (z)$ thus exhibits $N-2$ branch cuts, 
$\c C_p$, spanned between pairs of branch points $x_{2p-1}$ 
and $x_{2p}$, $p=1,\cdots ,N-2$, of $\omega(z)$, which 
correspond to zeros of the polynomial $Q_{2N-4}(z)$. 

Since the configuration of the Liouville charges at 
$w_p$ is one-dimensional, it is convenient to introduce the
linear density $\rho(w)$, defined when $w$ lies on $\c C$.
\begin{equation}
\rho (w) = { 1 \over 2\pi} \sqrt {R(w)-W'(w)^2 } 
\label{eq:rhodensity}
\end{equation}
The requirement that $\c C$ lie along branch cuts of $\omega$
does not determine the precise position of $\c C$. In fact, any
analytic curve that joins pairs of branch points would do. From
the fact that the Liouville charges at the points $w_p$ are all
of unit strength times $1/s$, it follows that the charge density
must be real and positive along $\c C$. This supplementary
condition requires that the position of the branch cut $\c C$,
supporting the Liouville charges at $w_p$, must be such that $dw
\rho(w)$ is real as $w$ is varied along $\c C$. When $\alpha$
and $\beta_i$ are real, this simply implies that the Liouville
charges at $w_p$ are concentrated on the real axis, as was
expected. However, the above conditions also provide consistent
prescriptions for the case when $\alpha$, $\beta _i$ as well as
the external momenta, are analytically continued to complex
values.

It remains to clarify the physical significance of 
the constants $R_i$, which enter the function 
$R(z)$ in (\ref{eq:rsol}) and the polynomial $Q_{2N-4}(z)$ 
in (\ref{eq:polyn}), in the $s \to \infty$ limit.
These relations are {\it automatically} satisfied 
by the construction of $\omega (z)$, as can be 
checked easily by taking the limit of (\ref{eq:omdef}) 
when $z\to z_i$. It thus appears that the $N-3$ 
independent parameters $R_i$, entering the 
solution of the second set of equations in 
(\ref{eq:elec})
are undetermined by these equations. 

How can this indeterminacy be understood ? It is easiest to
analyze first the case where all $z_i$, and all $a_i$ are
real. By construction, the $R_i$ are then real, as can be seen
from (\ref{eq:rsol}). When all $a_i$, $i=1,\cdots, N-1$ are
positive, (and only the compensating charge at $\infty$ is
negative) the possible locations for the Liouville charges are
on the $N-2$ line segments $\c C_p$, $p=1,\cdots , N-2$, in
between pairs of consecutive positive charges at $z_i$. However,
exactly how the total Liouville charge (which is fixed to be 1)
is partitioned among the $N-2$ line segments is not \`a priori
determined. Indeed, the positive Liouville charges cannot cross
over from one line segment into another, since crossing would
involve passing through a charge configuration of infinite
electrostatic energy when a Liouville charge is on top of a
charge $z_i$. Thus, for any partition of the Liouville charges
among the $N-2$ intervals, there must be an equilibrium
configuration, and the $N-3$ independent parameters $R_i$
precisely specify the possible partitions of the Liouville
charges over the $N-2$ intervals. When the points $z_i$ and the
charges $a_i$ are allowed to be complex, the line
segments on which the Liouville charges lie can move into the
complex plane,  but the counting is analogous.

In fact, the values of the parameters $R_i$ are 
determined by the first set of equations in 
(\ref{eq:elec}), which give the positions of the 
external vertex charges $z_i$, $i=1,\cdots, N-1$. 
In the high energy limit, where $s\to \infty$,  we have
\begin{equation}
\sum_{j=1\atop j\not=i}^{N-1}{2b_{ij}\over z_i-z_j} 
+{1 \over 4} R_i =0
\label{eq:firsteq}
\end{equation}
Therefore, out of the original $N-1$ equations in (\ref{eq:firsteq}), two
correspond to the asymptotic conditions on $\omega(z)$, leaving $N-3$
equations. In view of the analysis of the previous paragraph, only
$N-3$ real parameters amongst the $N-3$ complex $R_i$ are determined
by the electrostatics equations and the reality conditions, leaving
N-3 real parameters undetermined. 
\section{High Energy Limit from the Electrostatic Energy}
\noindent
The tree level correlation function, in the saddle point
approximation, is
\begin{equation}
    \corru{}{\prod_{i=1}^N \c V _i}    =  
    {\Gamma(-s) \mu ^s \over\alpha (4 \pi )^s}
      e^{-{\c E}_0}
      \label{eq:treefinal}
\end{equation}
where all the quantities are to be evaluated at the 
saddle point. In particular, the electrostatic 
equilibrium energy is given in terms of $\rho (w)$ 
and $W(w)$ by
\begin{equation}
    -{\c E _0\over2(\alpha s)^2}
   = 
    \half \sum_{{i,j=1\atop i\not=j}}^{N-1}   b_{ij} 
      \ln  |z_i-z_j |^2
     + \int _{\c C} dw \rho(w) \{ W(w)  + \bar W(w) \} 
    - \half \int _{\c C} dv\rho (v)\int _{\c C} 
   dw\rho (w)\ln |v-w|^2
   \label{eq:sdef}
\end{equation}
First, the saddle point equations for $\bar z_i$
and $\bar w_p$ are the same as for the quantities $ z_i$ and
$ w_p$, even when the charges $a_i$ and $b_{ij}$ are
complex. Second, the integration measure $dw\rho (w)$ must
be real along the line segments of charge density.
We see that, as a result, the entire electrostatic energy is a
sum of a contribution from $z_i$ and $w_p$ on the one hand, and
the same functional form, evaluated on $\bar z_i$ and $\bar w_p$
on the other hand. Thus, given the identity of the equations for
barred and unbarred quantities, the electrostatic energy is
just twice that evaluated on unbarred quantities only.
This simplified expression for the electrostatic energy at 
equilibrium can be recast in terms of the holomorphic potential  
\begin{equation}
  \Omega (z) = \int _{\c C} dw\rho(w)\ln(z-w)
\label{eq:phidef}
\end{equation}
of the Liouville charges at $w_p$. This potential is the 
analogue of the
potential $W$ for the charges $z_i$, and its derivative is
$\omega (z) = \Omega (z)'$. In terms of this function, we may
evaluate the electrostatic potential of the $w_p$-charges in a
simplified way:$^1$
 
\begin{equation}
  - {\c E _0\over2(\alpha s)^2} = \sum_{{i,j=1\atop
      i\not=j}}^{N-1} b_{ij} \ln (z_i-z_j ) - W_0 - \sum _
  {i=1}^{N-1} a_i \Omega (z_i)
\label{eq:simplify}
\end{equation}
These quantities are now all holomorphic, and as such will not
be changed upon continuous changes in the curve $\c C$. Thus,
any curve $\c C$, connecting the branch points can be used in
the expression above, which greatly simplifies its calculability.

In fact, it was further shown in [1] that the function $\Omega$ 
itself may be evaluated in terms of Abelian differentials. 
Let us simply quote the results :
\begin{equation}
\Omega (z) - \Omega (z_0) = - \half \sum _{i=1} ^{N-1}
 a_i \ln { z-z_i \over z_0 -z_i} - i \pi \int _{z_0} ^z 
 dw \rho (w)
 \label{eq:abelcom}
\end{equation}
Thus, the electrostatic energy at the saddle point is now
computed completely in terms of Abelian integrals on the
associated hyperelliptic surface.  The three and four point
functions may be worked out explicitly, and can be shown to exhibit 
exponential behavior in $u_{ij}$$^1$.


\begin{thebibliography}{000}
\bibitem{1}
K. Aoki and E. D'Hoker, ``Non-Critical Strings at High Energy'',
UCLA/96/TEP/27 preprint, hep-th/9609079
\bibitem{2}
E. D'Hoker, {\it Lecture Notes on 2-D Quantum Gravity 
and Liouville Theory}, in ``Particle Physics VI-th 
Jorge Andre Swieca Summer School'',
ed. O.J.P. Eboli, M. Gomes and A. Santoro, World 
Scientific Publishers, (1992);
P. Ginsparg and G. Moore, {\it Lectures on 2D Gravity and 2D String 
Theory}, TASI Lecture Notes, TASI summer School 1992, 
Published in Boulder
TASI 1992;
F. David, {\it Symplicial Quantum Gravity and Random Lattices}, 
hep-th/9303127;
``2-D Gravity in Non-Critical Strings'', Discrete and Continuum
Approaches, E. Abdalla, M.C.B. Abdalla, D. Dalmazi, A. Zadra, 
Springer-Verlag, 
Lecture Notes in Physics, {\bf 20}, 1994
\bibitem{3}
J.-L. Gervais and A. Neveu, Phys. Lett. {\bf B123} (1983) 86; 
Nucl. Phys. {\bf B224} (1983) 329;
J.-L. Gervais and A. Neveu, Nucl. Phys. {\bf B238 } (1984) 125; 
Comm. Math. Phys. {\bf 100} (1985) 15;
J.-L. Gervais and A. Neveu, Phys. Lett. {\bf B151} (1985) 271; 
Nucl. Phys. {\bf B264} (1986) 557;
A. Bilal and J.-L. Gervais, Phys. Lett. {\bf B187 } (1987) 39; 
Comm. Math. Phys. {\bf 100} (1985) 15
\bibitem{4}
D.J. Gross and P. Mende, Phys. Lett. {\bf B197} (1987) 129,
Nucl. Phys. {\bf B303 } (1988) 407;
D.J. Gross, Phys. Rev. Lett. {\bf 60} (1988) 1229;
D. Amati, M. Ciafaloni, G. Veneziano, Int. J. Mod. Phys. {\bf A3} (1988) 1615, 
Phys. Lett. {\bf 197B} (1987) 81
\bibitem{5}
M. Goulian and M. Li, Phys. Rev. Lett. {\bf 66} (1991) 2051;  
K. Aoki and E. D'Hoker, Mod. Phys. Lett. {\bf A3} (1992) 235
\bibitem{6}
Vl.S. Dotsenko and V.A. Fateev, Nucl. Phys. {\bf B240} (1984) 312,
{\bf B251} (1985) 691
\end{thebibliography}
\end{document}